\documentclass[conference]{IEEEtran}
\IEEEoverridecommandlockouts
\usepackage{cite}
\usepackage{amsmath,amssymb,amsfonts}
\usepackage{caption}
\usepackage{subcaption}
\usepackage{algorithm}
\usepackage{algorithmic}
\usepackage{amsmath}
\usepackage{graphicx}
\usepackage{textcomp}
\usepackage{xcolor}
\usepackage{orcidlink}
\def\BibTeX{{\rm B\kern-.05em{\sc i\kern-.025em b}\kern-.08em
    T\kern-.1667em\lower.7ex\hbox{E}\kern-.125emX}}
    
\newcommand{\INDSTATE}[1][1]{\STATE\hspace{#1\algorithmicindent}}

\renewcommand{\baselinestretch}{0.97}

\begin{document}

\title{Fibbinary-Based Compression and Quantization for Efficient Neural Radio Receivers\\}

\author{Roberta Fiandaca \textsuperscript{\orcidlink{https://orcid.org/0009-0008-9665-4135}}, Manil Dev Gomony \textsuperscript{\orcidlink{https://orcid.org/0000-0002-5889-0785}}, \textit{Member, IEEE}}

\maketitle

\begin{abstract}
Neural receivers have shown outstanding performance compared to the conventional ones but this comes with a high network complexity leading to a heavy computational cost. This poses significant challenges in their deployment on hardware-constrained devices.
To address the issue, this paper explores two optimization strategies: \textit{quantization} and \textit{compression}.
We introduce both \textit{uniform}
and \textit{non-uniform} quantization
such as the \textit{Fibonacci Code word Quantization} (FCQ).
A novel \textit{fine-grained} approach to the \textit{Incremental Network Quantization} (INQ) strategy is then proposed to compensate for the losses introduced by the above mentioned quantization techniques. Additionally, we introduce two novel \textit{lossless compression} algorithms that effectively reduce the memory size by compressing sequences of Fibonacci quantized parameters characterized by a huge redundancy. The quantization technique provides a saving of 45\% and 44\% in the multiplier's power and area, respectively, and its combination with the compression determines a 63.4\% reduction in memory footprint,
while still providing higher performances than a conventional receiver.
\end{abstract}


\section{Introduction}
Wireless technology has advanced rapidly, with each generation introducing significant improvements: 5G, for example, delivers higher bandwidth and lower latency than its predecessors. However, the growing demand for faster data rates and greater network capacity is driving research toward sixth-generation (6G) wireless networks. Meeting the stringent requirements of 6G requires the integration of emerging technologies such as Artificial Intelligence (AI), which excels at solving complex computational problems. However, neural networks remain resource intensive, posing challenges for deployment on devices with limited hardware and strict latency constraints. To address this, optimization techniques at both the algorithmic and architectural levels can reduce power consumption and silicon area, but they must be applied carefully to preserve system performance. 

In this work, we propose novel techniques to reduce power dissipation and area footprint in an AI-based 6G physical layer architecture. Our reference model is a neural receiver capable of outperforming conventional receivers and approaching the performance of ideal ones. However, this network requires an extremely large number of multiplications and parameters, resulting in excessive power and area consumption. To address this issue, which remains largely unexplored in the literature, we investigate two optimization approaches: \textit{quantization} and \textit{compression}. Concerning the former, we mainly adopt the aggressive Fibonacci Codeword Quantization (FCQ) technique \cite{Fibonacci_paper}. It is combined with \textit{Incremental Network Quantization (INQ)} \cite{Fibonacci_paper}, which preserves high accuracy under aggressive quantization by progressively quantizing the network. We further propose an enhanced strategy that introduces \textit{finer-grained} quantization steps whenever performance degradation is detected. This approach allows a greater number of layers to be quantized using the Fibonacci scheme while maintaining acceptable performance levels.

For compression, we develop two novel lossless algorithms applied sequentially: \textit{word-length compression} and \textit{word-count compression}. The former is based on the Zeckendorf theorem \cite{Zeckendorf} and reduces the number of bits required to represent each weight. The latter operates on the word-length–compressed data and targets the number of weights to be stored, leveraging the significant redundancy introduced by FCQ. The combined use of the proposed quantization and compression techniques enables substantial reductions in memory usage, as well as in the area and power consumption of the multipliers, and consequently of the entire network, while maintaining satisfactory performance. This effectively fulfills our objective of reducing the overall complexity of the network.

In what follows, we shall provide a brief overview of (i) the chosen neural receiver architecture, (ii) the quantization technique, and (iii) the compression (Section \ref{Sec:background}); explain the developed quantization (Section \ref{sec:Quantization}) and compression (Section \ref{sec:Compression}) techniques before concluding with some final remarks and perspective of this work (Section \ref{sec:Conclusion}).

\section{Background information}
This section will first present the neural receiver under investigation, followed by an overview on neural network quantization and an introduction to compression.
\label{Sec:background}
\subsection{Neural Receiver}
\label{sec:Neural Receiver}
The reference architecture to be optimized is the neural receiver proposed in the software framework Sionna, 
a TensorFlow-based open-source library for Next-Generation Physical Layer Research \cite{sionna}.
The architecture is a Single-Input Multiple-Output (SIMO) baseband communication system characterized by a single antenna at the transmitter and multiple antennas at the receiver side \cite{DeepRx}. The neural receiver replaces the channel estimator, the equalizer and the demapper. This receiver is based on the DeepRx \cite{DeepRx} which produces better performances than the traditional receiver as it learns multiple blocks simultaneously (i.e., it performs the channel estimation, the equalization and the demapping tasks jointly), rather than optimizing each block individually.
The performances are expressed in terms of Bit Error Rate (BER) against the normalized Signal-to-Noise Ratio ($E_b/N_0$).
A comparison between the neural receiver and the two baselines shows the good performances of the former: they are, in fact, very close to the ideal receiver, based on full channel knowledge, outperforming the traditional one, based on the Least-Square (LS) Estimation instead (see Figure \ref{fig:16bits_8bits_PTQ}).

\subsection{Neural Network Quantization} 
\label{Sec:Neural Network Quantization}
Quantization can be categorized into uniform methods, characterized by equally spaced quantization steps, and non-uniform methods, where the step sizes vary. In this work, we adopt the Fibonacci Codeword Quantization (FCQ)~\cite{Fibonacci_paper}. FCQ employs non-constant step sizes, making it a non-uniform quantization scheme. Unlike traditional non-uniform techniques, FCQ is designed primarily to reduce the area and power consumption of multipliers in the network, rather than to improve numerical precision. In the following, we refer to FCQ as an aggressive quantization technique. FCQ operates by rounding each number to its nearest Fibonacci codeword, i.e., a value whose binary representation contains no consecutive ones. However, when applied in isolation, this aggressive quantization leads to significant accuracy degradation. To mitigate this issue, we employ Incremental Network Quantization (INQ) \cite{Fibonacci_paper}, which incrementally quantizes subsets of weights, freezes their updated values, converts the model back to floating-point representation, and retrains the remaining portions of the network. To measure the noise introduced by the quantization we shall employ the \textit{Mean-Squared Error} (MSE). The lower the MSE, the closer the performances will be to the non-quantized network ones.


\section{Non-uniform Fibonacci Quantization}
\label{sec:Quantization}

The outstanding performances of the neural receiver presented in the Section \ref{sec:Neural Receiver} come with a significant complexity as they require a huge number of multiplications due mainly to the convolutional layers and to a large number of parameters. Specifically, considering the ten convolutional layers (Conv2D) and the eight normalization layers, we estimate a total number of multiplications equal to $1.6 \cdot 10^{11}$ and $3.4 \cdot 10^6$ parameters by summing up each layer contribution.
This motivates us to explore and implement methods for optimizing the network, i.e., to apply techniques in order to reduce the area and power consumption making the hardware deployment possible.

\begin{figure*}
	 \centering
	\begin{minipage}{1\columnwidth}
		\centering
		\includegraphics[width=0.75\columnwidth]{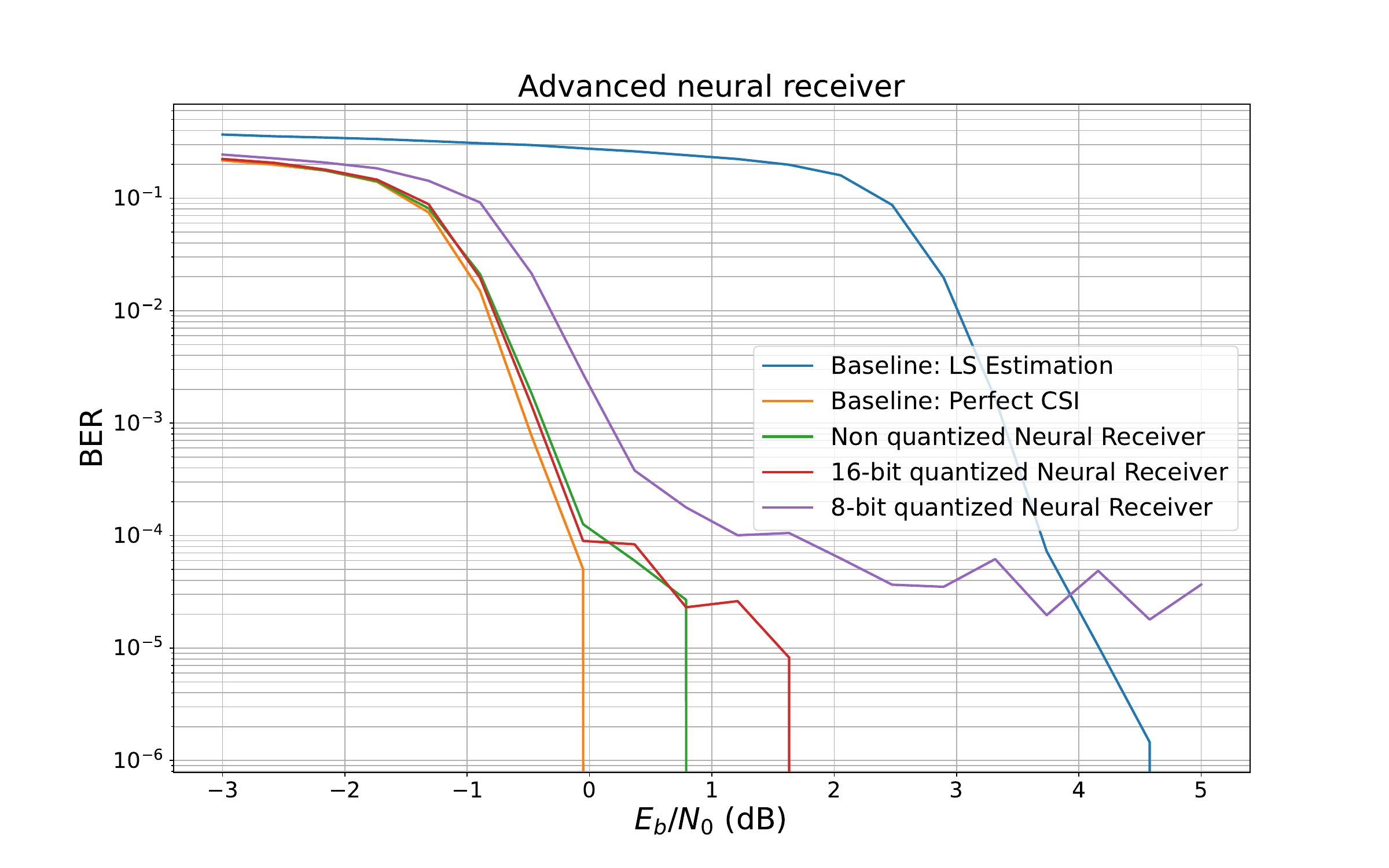}
		\subcaption{}
		\label{fig:16bits_8bits_PTQ}
	\end{minipage}
	\begin{minipage}{1\columnwidth}
		\centering
		\includegraphics[width=0.75\columnwidth]{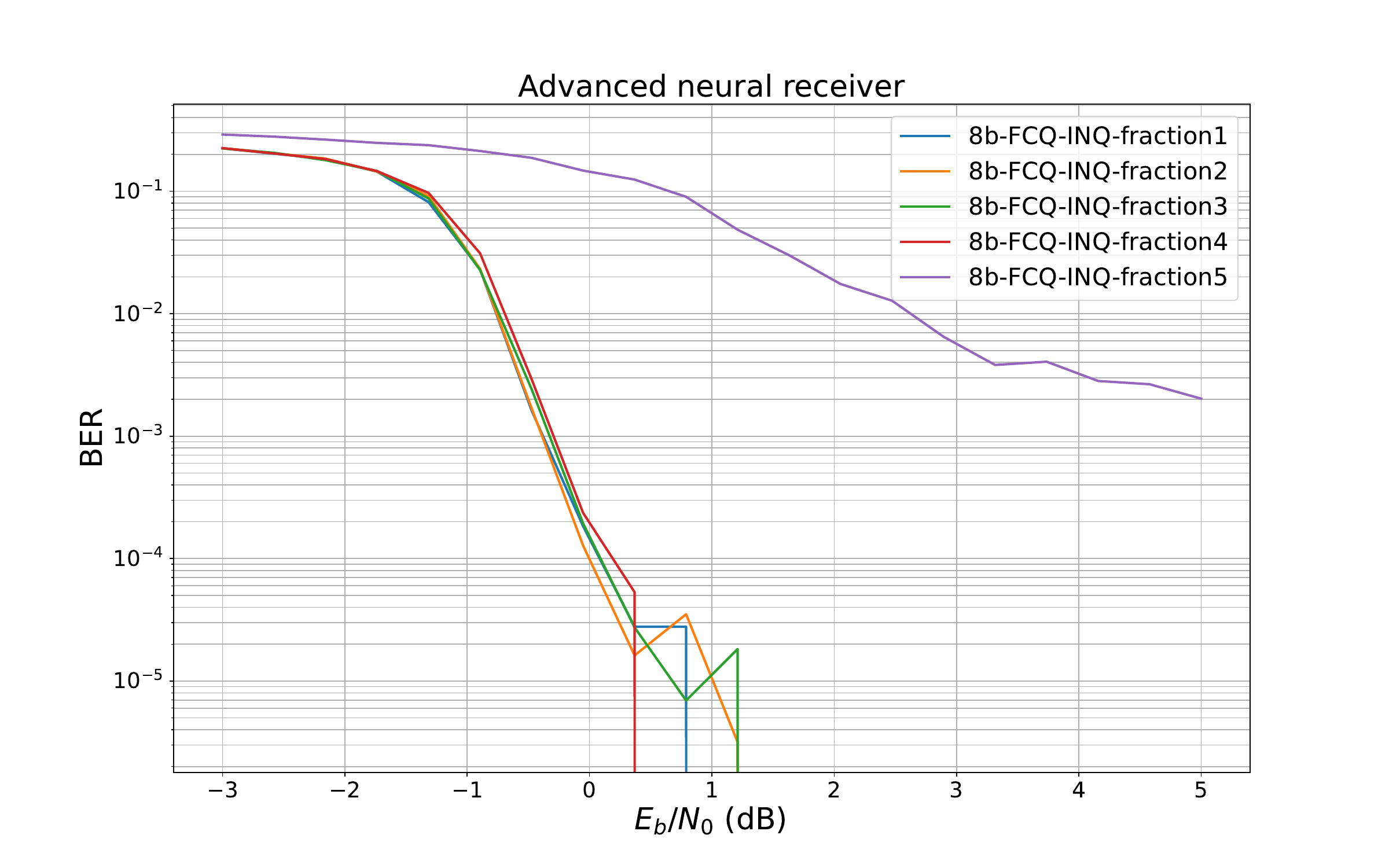}
        \subcaption{}	\label{fig:fractions}
       \centering
      \end{minipage}%
    \caption{\small(a) Performance comparison (BER against $E_b/N_0$) between LS-based receiver, perfect CSI receiver, non-quantized neural receiver, 16-bit quantized neural receiver, 8-bit quantized neural receiver; (b) 8-bit Fibonacci quantization with INQ strategy: \textit{fraction} approach. We recall that the performances of the non-quantized and the 16-bit quantized networks are equivalent.}
    \vspace{-4mm}
\end{figure*}

Quantization is the first technique adopted with the purpose of reducing the occupied memory and the computational requirements of the neural receiver. The one
adopted in our simulations is a \textit{post-training} quantization, i.e., the quantization is performed over the already-trained weights. Let us now deepen the application of this technique to our architecture. 
\\
    \textbf{Uniform quantization.}
    The uniform quantization is applied as a first step to the neural receiver optimization: the 32-bit floating point (FP32) parameters are transformed into numbers of type quint16 and quint8, see Figure \ref{fig:16bits_8bits_PTQ}). 

\begin{figure*}
	
     \centering
	\begin{minipage}{1\columnwidth}
		\centering
		\includegraphics[width=0.75\columnwidth]{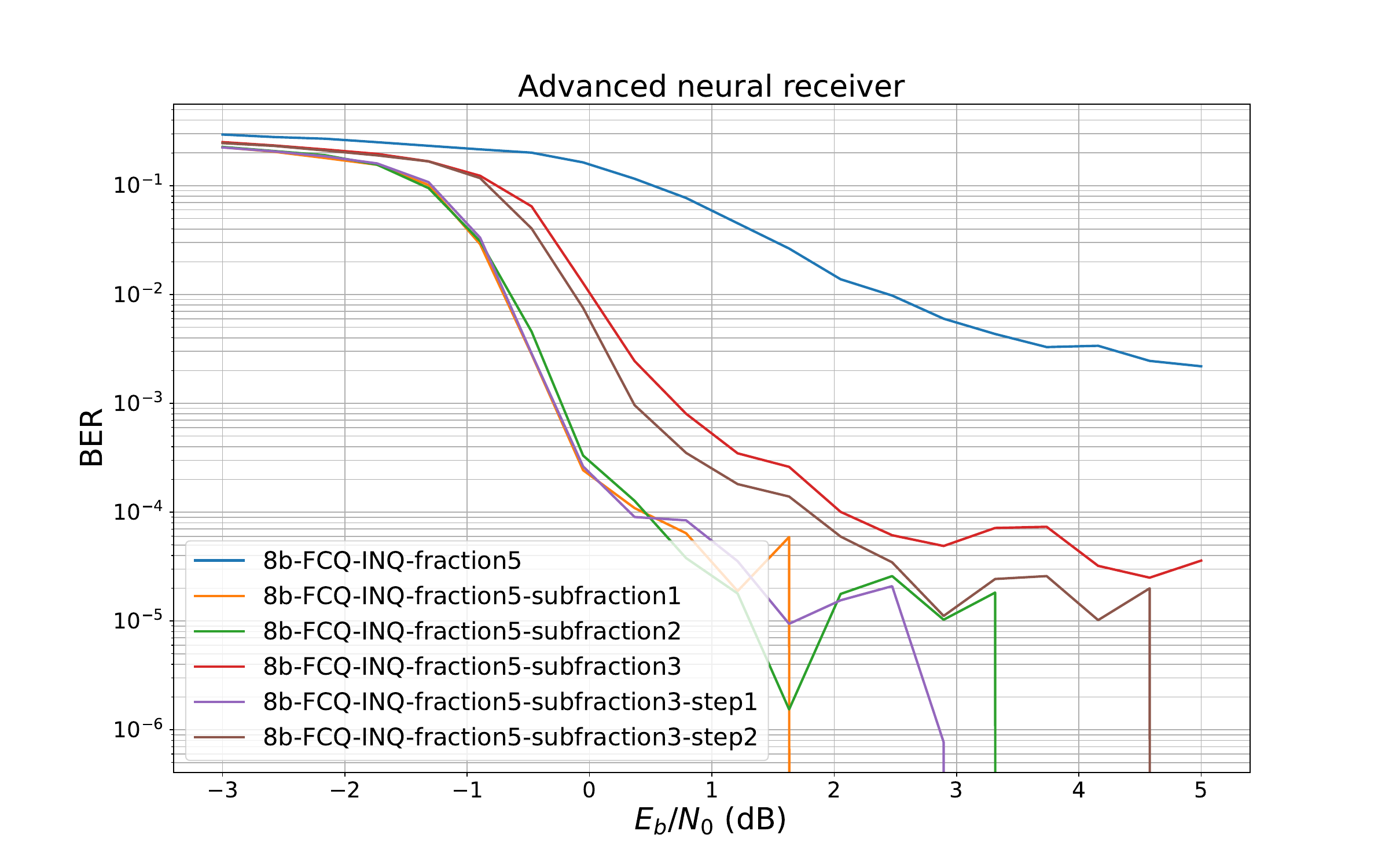}
        \subcaption{}
        \label{fig:subfractions}
	\end{minipage}%
    \centering
	\begin{minipage}{1\columnwidth}
		\centering
		\includegraphics[width=0.75\columnwidth]{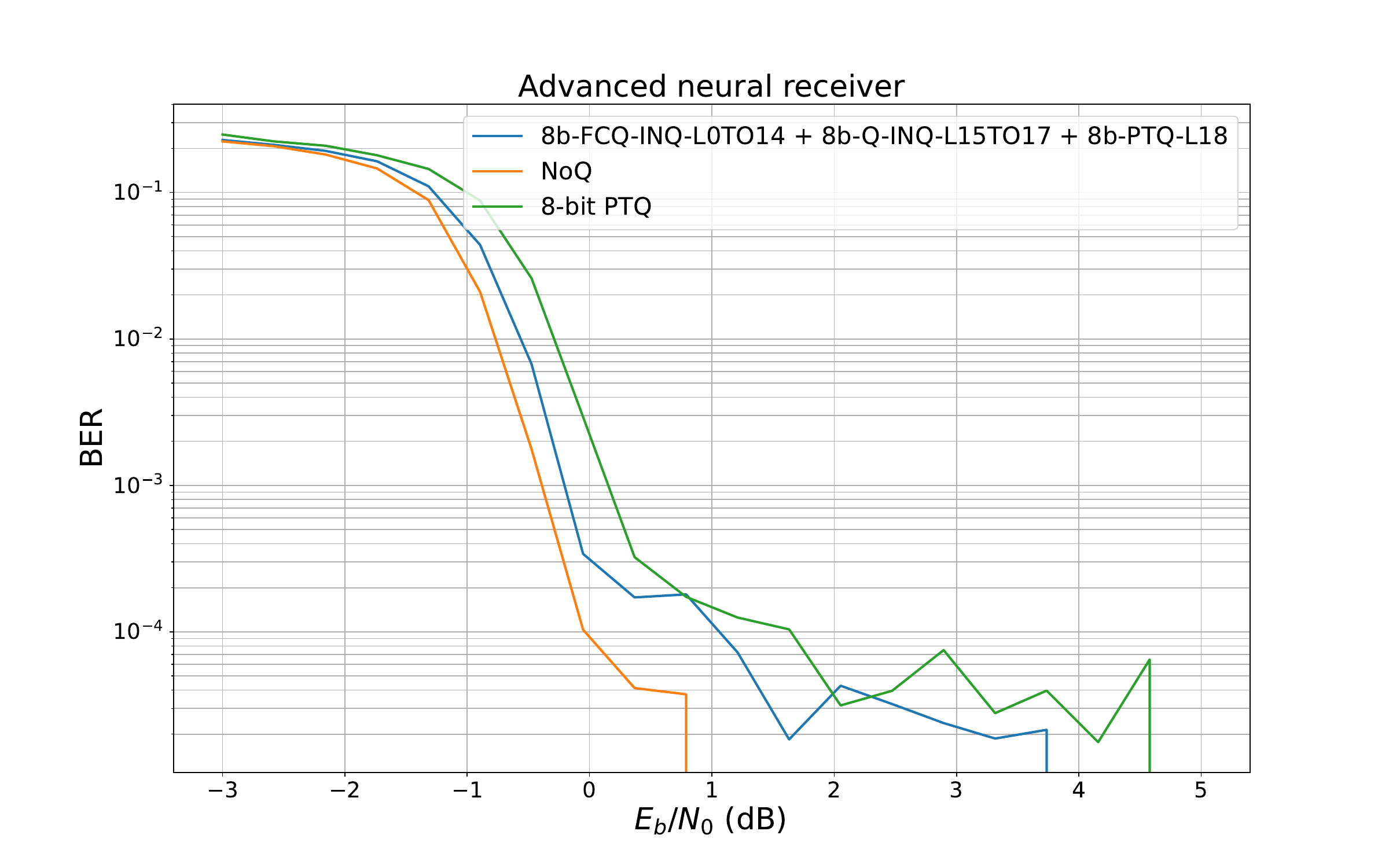}
        \subcaption{}
		\label{fig:Final}
	\end{minipage}%
    \caption{\small (a) 8-bit Fibonacci quantization with INQ strategy: \textit{more fine-grained } approach for fraction 5; (b) Performance comparison between the non-quantized network (NoQ), the 8-bit quantized network (8b-PTQ) and the optimized network (8b-FCQ-INQ-L0TO14 + 8b-Q-INQ-L15TO17 + 8b-PTQ-L18).}
    \vspace{-4mm}
\end{figure*}

The 16-bit quantized neural receiver performs almost exactly as the non-quantized one. This is confirmed by the value of the MSE which is of the order $10^{-9}$. 
The 8-bit quantization, instead, introduces an evident performance degradation, still performing better than the traditional receiver.
The advantage offered by the 16-bit and 8-bit integer quantizations is a reduction of a factor 2x and 4x in model size and memory bandwidth requirements, respectively.
Moreover, the integer quantization allows for a faster hardware computation compared to FP32.
\\
    \textbf{Non-Uniform quantization.}
    We apply different quantization techniques to different layers of the structure in order to improve the neural receiver accuracy. Specifically, we adopt an \textit{aggressive} quantization for the first layers and a \textit{uniform} quantization for the last layers. The aggressive quantization is the so-called \textit{Fibonacci Code word Quantization} (FCQ), which is mitigated via the \textit{Incremental Network Quantization} (INQ) strategy, as described in Section \ref{Sec:Neural Network Quantization}. Differently from \cite{Fibonacci_paper}, where the two techniques are used to quantize the weights of the network, we shall apply them to \textit{tensors} of weights. 
The smaller the set of tensors to quantize, the better the accuracy gain. On the other hand, the long training time prohibits to use very small sets. 
We stress here that the FCQ is applied only to the weights' tensors; this is because the biases are not involved in the multiplication. Therefore, it is not beneficial to apply them the FCQ strategy which focuses on reducing multiplier's area and power. 
To achieve the core idea of incrementality we choose to quantize an \textit{increasing} number of tensors at each INQ iteration. In particular, we quantize one layer at the first INQ fraction, two layers at the second fraction and so on to the fifth fraction, where we quantize five layers. The remaining three tensors are Fibonacci encoded at the sixth fraction. Looking at Figure \ref{fig:fractions}, we can notice a significant performance degradation at the fifth fraction.
For this reason, we introduce a novel approach consisting in the use of \textit{subfractions}. The application of a more fine-grained approach greatly reduces the accuracy degradation as it introduces a lower perturbation in the network. Its combination with the retraining operation, which further reduces the loss, makes such an accuracy degradation minimal. Using this method, the fifth fraction is further divided into subfractions of $1 + 2 + 2$ tensors: at the third of these iterations, it is possible to observe, again, a performance degradation (see Figure \ref{fig:subfractions}).
This leads to the introduction of two more steps, quantizing one tensor each (Figure \ref{fig:subfractions}).
\\
The performances become worse than the traditional 8-bit quantization starting from layer 14. 
The remaining four layers, which are still in floating point, are optimized through a uniform 8-bit quantization in combination with the INQ strategy, necessary to recover part of the accuracy lost during the quantization. Obviously, concerning the last layer, since all the others are frozen at this stage, the INQ strategy reduces to a simple PTQ approach. The achieved performances are shown in Figure \ref{fig:Final}, compared to those of the 16-bit (equal to the non-quantized) and 8-bit quantized networks. 

The introduction of the FCQ has allowed the usage of OR-based approximate multiplier reducing both the area and power consumptions compared to the standard multiplier. Specifically, $\frac{n^2-n}{2}$ out of the original  $(n-2)\cdot n$ FAs in an $n$-bit carry save multiplier are replaced by OR gates \cite{Fibonacci_paper}, making the savings in our 8-bit scenario equal to $58\%$. Furthermore, the area occupied by an OR gate is four times smaller than the FA's one. Thus, the total area saving obtained by using an approximate multiplier is $44\%$.
The introduction of the FCQ technique is beneficial also to the power consumption which is reduced of an amount equal to $45\%$, considering that the OR gate consumes  $77\%$ less power with respect to a FA.

 Other kinds of \textit{non-uniform quantization} techniques, such as the Power-of-Two \cite{Po2} or the dynamic quantization \cite{dynamic_quantization}, can further improve the performance-complexity trade-off. Moreover, replacing PTQ with QAT techniques could yield  better results. Finally, since the architecture is characterized by a noisy channel that deeply impacts the performances, a possible direction could be the adoption of a \textit{dynamic precision}, higher for lower $E_b/N_0$ values.

\section{Fibbinary Compression}
\label{sec:Compression}
Despite the application of quantization, the network still requires a large number of parameters to be stored, making it memory intensive and challenging to deploy on systems with limited hardware resources.
To overcome this issue, we develop a \textit{lossless} compression method consisting of two algorithms: the \textit{word-length} compression, which reduces the number of bits for each weight, and the \textit{word-count} compression, which diminishes the total number of weights. 
The compression is applied on top of the 8-bit Fibonacci quantized weights and the two algorithms are applied sequentially. Let us now delve into each of the two algorithms.
\\
\textbf{Word-length compression} The word-length compression is achieved by exploiting some properties of the \textit{Fibbinary} numbers: these are the values that we have referred to as ``Fibonacci quantized numbers" in Section \ref{sec:Quantization}, according to the nomenclature in \cite{Fibonacci_paper}, i.e. those number that does not contain two consecutive ones in their binary representation. The Fibbinary numbers are related to the Fibonacci numbers through the Zeckendorf expansion of integer numbers \cite{Zeckendorf}, according to which every positive integer $n$ can be written uniquely as the sum of distinct, non-consecutive Fibonacci numbers:

\begin{equation}
   n = F(k_1) + F(k_2) + .... + F(k_m) 
\end{equation}
where $k_i$ is the index of the Fibonacci number $F(k_i)$.
The $n^{th}$ Fibbinary number is:
\begin{equation}
   a_n = 2^{k_1-2} + 2^{k_2-2} + ... + 2^{k_m-2}.
\end{equation}
By exploiting the relationship between Fibbinary and Fibonacci numbers, it is no longer necessary to store the Fibbinary values of the weights but it is sufficient to save their respective indices. In our scenario, we aim
to compress 8-bit Fibbinary weights. 
With 55 possible Fibbinary numbers that can be generated using 8 bits, we only require 6 bits to store the indices of the weights instead of the 8 bits needed to store the weights themselves.
\\
\textbf{Word-count compression}
The second step for the compression consists in reducing the total number of weights by exploiting the huge redundancy shown by the Fibbinary weights. Since they are not consecutive, it is not possible to use already existing compression algorithms applied to remove redundancy when repetitive data are consecutive (e.g., run-length encoding).
To address this issue, we develop a \textit{novel algorithm} to compress series of redundant \textit{consecutive} and \textit{non-consecutive} values.
This algorithm is applied on top of the previously word-length compressed weights, represented using 6 bits, and introduces an overhead of two bits. Therefore, a total number of 8 bits is required for storage after the word-count compression.
The key idea of the algorithm is to detect the two most common numbers appearing in each tensor of weights (by reading one time the sequence) and store them only once in the memory, thus avoiding to save them every time they are encountered.
The procedure is shown in Algorithm \ref{word-count compression}.

\begin{algorithm}
  \algsetup{linenosize=\small}
\caption{Word-count Compression}\label{word-count compression}
\begin{algorithmic}
\STATE Count each number's repetitions
\STATE {Identify the two most common numbers}
\INDSTATE $\textit{A} \gets \text{most common number}$
\INDSTATE $\textit{B} \gets \text{second most common number}$
\INDSTATE $\textit{C} \gets \text{other numbers}$
\STATE {Sequentially readout the numbers and:} \\
\textbf{If} {$A+C$:} \\
  \INDSTATE Store ``01" + \text{$C$;} \\
\textbf{If} {$B+C$:} \\
\INDSTATE Store ``10" + \text{$C$;} \\
\textbf{If} {$C+C_1\ |\ C_1 \neq A,B$:} \\
\INDSTATE Store ``00" + \text{$C$;} \\
\INDSTATE Store ``00" + \text{$C_1$;} \\
\textbf{If} {$n\cdot A$ \textsc{AND} $m\cdot B\ |\ n,m \in \mathbb{N}_0$:} 
\\
\INDSTATE[1] $\text{Store}\ ``11"\ +\ \text{bit}_{3..8}$; \\
\INDSTATE[2] \textbf{If} {Last number read in the sequence $=\ A$:}\\
\INDSTATE[3] $\text{bit}_3 = ``0"$; \\
\INDSTATE[2] \textbf{Else if} {Last number read in the sequence $=\ B$:} \\
\INDSTATE[3] $\text{bit}_3 = ``1"$; \\
\INDSTATE[2] \textbf{If} {$n \neq 0$ \textsc{AND} $m \neq 0$:} 
\INDSTATE[3] $\text{bit}_4=``1"$; $\text{bit}_{5,6,7} = n$; $\text{bit}_{8}= m$; \\
\INDSTATE[2] \textbf{Else} 
\INDSTATE[3] $\text{bit}_4=``0"$;  $\text{bit}_{5...8} =n (m=0), m (n=0)$;
\end{algorithmic}
\vspace{-1mm}
\end{algorithm}The codeword is stored when a $C$ is read or when the consecutive repetitions exceed the one representable on 4 bits.
When this latter scenario occurs, the sequence is further divided into several pieces, and each of them undergoes the previously explained rules.
\begin{figure}
    \centering
    \includegraphics[width =\columnwidth ]{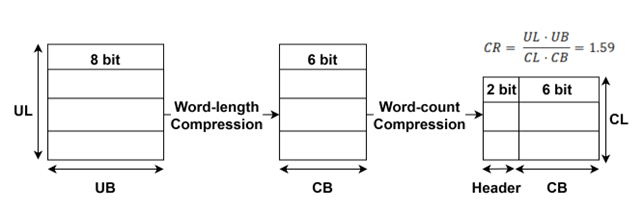}
    \caption{\small Pictorial representation of the word-length and word-count compressions algorithms. UL stands for \textit{Uncompressed Length}, i.e. the number of weights in the tensors before the compression. UB describes the \textit{Uncompressed Bitwidth}. CB represents \textit{Compressed Bithwidth}. CL is the \textit{Compressed Length}, i.e. the tensors' length after the compression. The compression ratio (CR) is computed as in the picture.}
    \label{fig:CR}
    \vspace{-4mm}
\end{figure}

The savings achieved through the introduction of compression are expressed in terms of Compression Ratio (CR), that we find to be equal to $1.59$. This number can be explained by looking at the Figure \ref{fig:CR}, considering that $UL=1843840$ and $CL=1160744$. Moreover, we found a variation to the algorithm that provides a slightly better result in terms
of CR. It consists in applying the compression to a group of tensors rather than to the single ones. This means that the two most common numbers are searched among $k$
tensors grouped together. The optimal result, following this strategy, is achieved
by compressing three tensors at a time ($k = 3$) and the resulting CR is found to be $1.63$. We note that the compression technique, like quantization, is applied to the weights' tensors only, as it operates on top of Fibbinary numbers. Thus, it is not applied to the biases which are not considered in the calculation of the CR values reported above.
Going to possible future perspectives, the development of lossy compression techniques would achieve a better compression ratio (guaranteeing, on the other side, as lower distortion as possible). Furthermore, the use of neural network compression strategies, like \textit{pruning} \cite{Pruning} or \textit{knowledge distillation} \cite{knwoledge_distillation} would reduce the size of the model.

\section{Conclusions}
\label{sec:Conclusion}
In this work, we presented efficient quantization and compression techniques that allowed us to obtain a total (considering the entire structure, thus tensors of both weights and biases) \textit{memory area saving} equal to $63.8\%$ compared to the 16-bit quantization and equal to $26\%$ compared to the 8-bit quantization.
In the first case, if on one hand we have slightly degraded performance, on the other hand we reach a significant amount of area reduction; in the second case, instead, the reduction in memory space is lower but we gain in performance, as shown in Figure \ref{fig:Final}.  
This remarkable network optimization makes it more accessible for widespread deployment in a variety of applications. 
The techniques introduced and developed in this work will play a crucial role in meeting the stringent requirements in terms of area, power consumption, and latency of AI-based receivers required in future wireless communications systems. 

\bibliographystyle{abbrv}
\bibliography{bibliography.bib}

\end{document}